\documentclass{PoS}
\usepackage{epsfig}
\usepackage{cite}

\newcommand{\bc}{\begin{center}}
\newcommand{\ec}{\end{center}}
\newcommand{\be}{\begin{equation}}
\newcommand{\ee}{\end{equation}}
\newcommand{\bea}{\begin{eqnarray}}
\newcommand{\eea}{\end{eqnarray}}
\newcommand{\ba}{\begin{array}}
\newcommand{\ea}{\end{array}}
\newcommand{\lb}{\label}
\newcommand{\rf}{\ref}
\newcommand{\bfg}{\begin{figure}[htbp]}
\newcommand{\efg}{\end{figure}}

\newcommand{\prd}{Phys. Rev. D }
\newcommand{\np}{Nucl. Phys. }
\newcommand{\npb}{Nucl. Phys. B }
\newcommand{\prl}{Phys. Rev. Lett. }
\newcommand{\prp}{Phys. Rep. }

\newcommand{\plb}{Phys. Lett. B }

\newcommand{\rmp}{Rev. Mod. Phys. }

\newcommand{\ptep}{Prog. Theor. Exp. Phys. }
\newcommand{\zp}{Z. Phys. }

\newcommand{\epjc}{Eur. Phys. J. C }
\newcommand{\jhep}{JHEP }

\title{Large-${N_{\mathrm{c}}^{}}$ QCD and tetraquarks}

\ShortTitle{Large-${N_{\mathrm{c}}^{}}$ QCD and tetraquarks}

\author{\speaker{Hagop Sazdjian}\\
IPNO, Universit\'e Paris-Sud, CNRS-IN2P3, Universit\'e Paris-Saclay, 
91405 Orsay, France\\
E-mail: \email{sazdjian@ipno.in2p3.fr}}

\abstract{Tetraquark properties are examined in the limit of large
${N_{\mathrm{c}}^{}}$ of color in QCD. The qualitative differences
between molecular and compact tetraquarks are outlined. Consequences
of the possible existence of compact tetraquarks are analyzed and
shown to lead to upper bounds in the $N_{\mathrm{c}}^{}$-behavior of
their decay widths. Open questions on theoretical grounds, related
to the dynamics of systems composed of two quarks and two antiquarks,
are addressed.}

\FullConference{XIII Quark Confinement and the Hadron Spectrum -
Confinement2018\\
31 July - 6 August 2018\\
Maynooth University, Ireland}

\begin{document}

\section{Multiquark states in QCD} \lb{s1}

Hadrons are color-singlet bound states of quarks and gluons. Mesons
are essentially made of a quark and and an antiquark (plus gluons and
sea quarks), while baryons are essentially made of three quarks 
(plus gluons and sea quarks). A schematic representation of mesons
and baryons is given in Fig. \rf{f1}.
\par
\bfg 
\bc
\epsfig{file=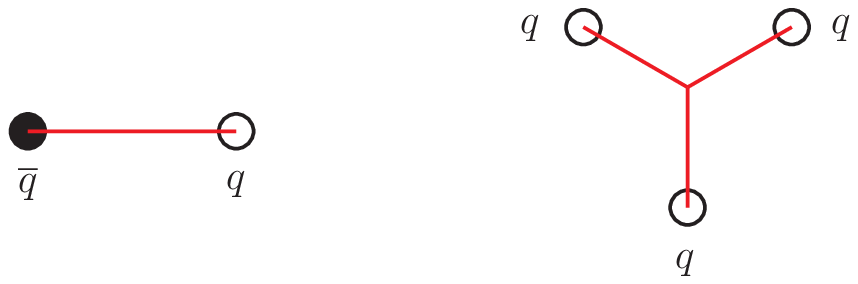,scale=0.75}
\caption{Schematic representation of mesons and baryons.
The straight lines represent gauge links.} 
\lb{f1} 
\ec
\efg
\par
Are there other types of structure for bound states in QCD,
which might be classified as exotics with respect to the
conventional hadronic states? In principle, this would be
possible by constructing gauge invariant composite operators
which might generate exotic states. Examples of such states
are tetraquarks, made essentially of two quarks and two
antiquarks, and pentaquarks, made essentially of four quarks
and one antiquark (Fig. \rf{f2}).
\par
\bfg 
\bc
\epsfig{file=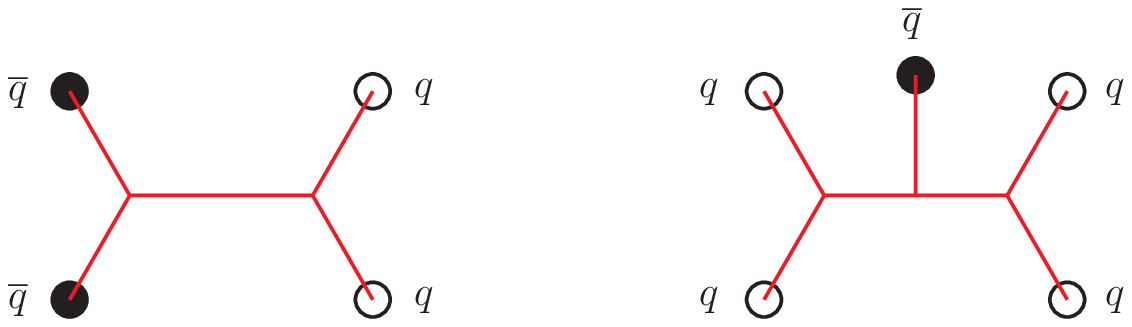,scale=0.75}
\caption{Schematic representation of tetraquarks and pentaquarks.
The straight lines represent gauge links.} 
\lb{f2} 
\ec
\efg
\par
The possibility of the existence of multiquark states has been
considered by many authors and in particular by Jaffe
in the framework of the bag model \cite{Jaffe:1976ig}. 
However, theoretical difficulties arise in QCD. We concentrate
in the following on the tetraquark problem. The difficulty is related
to the fact that a tetraquark field, local or nonlocal, made of a
pair of quark and antiquark fields, which would be color-gauge
invariant, can be decomposed, by Fierz transformations, into a
combination of products of color-singlet bilinear operators of
quark-antiquark pairs. For instance, in local form, a four-quark
color-singlet operator $T(x)=(\overline q\overline qqq)(x)$ can
be decomposed as
\be \lb{e1}
T(x)\equiv (\overline q\overline qqq)(x)\sim
\sum (\overline qq)(x)(\overline qq)(x),
\ee
where the $(\overline qq)(x)$s are themselves color-singlet
operators.
\par
However, color-singlet bilinears esssentially describe ordinary
meson fields or states. The above decomposition is suggestive of
a property that tetraquarks would be factorizable into
independent mesons and could at best be bound states or resonances
of mesons \cite{Voloshin:1976ap,Bander:1975fb,DeRujula:1976zlg,
Tornqvist:1993ng,Amsler:2004ps,Swanson:2006st}, also called
\textit{molecular tetraquarks}, and not genuine bound states of
two quarks and two antiquarks, which would result from the direct
confinement of the four constituents.
\par
What would be, on phenomenological grounds, the difference of the
two types of bound state, since both of them would be represented
by poles in the hadronic sector?
\par
Meson-meson interaction forces are short-range and weak, as
compared to the strong long-distance confining forces. Therefore,
molecular type tetraquarks would be loosely bound states, with
relatively large space extensions, while tetraquarks formed
directly by confining forces would be more tightly bound. The
latter are also called \textit{compact tetraquarks}. Compact
tetraquarks would also exist in flavor multiplicities, since
confinement is independent of flavor.
\par
The above qualitative differences have their influence on
phenomenological quantities, like the number of states, decay
modes, decay widths and transition amplitudes.
\par
For more than ten years, many tetraquark candidate states have been
signalled by several experiments: BaBar, Belle, BESIII, CDF, CLEO,
D0, ATLAS, CMS, LHCb. Ordinary meson structures could not fit their
properties. An intense theoretical activity has been developed around
the extraction of their properties and their interpretation. However,
there are difficulties to explain all data by a single model or
mechanism. Many review articles give thorough descriptions of the
various facets of the problem \cite{Chen:2016qju,Hosaka:2016pey,
Esposito:2016noz,Lebed:2016hpi,Guo:2017jvc,Ali:2017jda,Olsen:2017bmm,
Karliner:2017qhf}.
\par
In lattice calculations, a general consensus does not yet seem
to exist, a majority of investigations not providing
evidence for tetraquarks \cite{Ikeda:2013vwa,Padmanath:2015era,
Ikeda:2016zwx,Cheung:2017tnt,Hughes:2017xie}, while evidence is found
in a few sectors involving the $b$ quark \cite{Francis:2016hui,
Bicudo:2016ooe}. 
\par
We shall be interested here by the qualitative properties of compact
tetraquarks, since the very existence of compact tetraquarks is
intimately related to fundamental properties of QCD, not yet well
understood. The existence of molecular type tetraquarks does not
raise any conceptual difficulty.
\par
The \textit{diquark model} has been proposed to explain the formation
of compact multiquark states \cite{Schafer:1993ra,Jaffe:2003sg,
Nussinov:2003ex,Shuryak:2003zi}. It has been applied to tetraquark
phenomenology by Maiani, Piccinini, Polosa and Riquer
\cite{Maiani:2004vq,Maiani:2017kyi}. Other approaches
with the diquark mechanism can be found in
\cite{Ebert:2007rn,Heupel:2012ua,Brodsky:2014xia,Lebed:2017min}.
\par
We shall study the tetraquark problem in the
large-$N_{\mathrm{c}}^{}$ limit of QCD, which might give us
complementary informations about it.
\par

\section{QCD at large {\boldmath{$N_{\mathrm{c}}^{}$}}} \lb{s2}

The framework that is considered is that of
$\mathrm{SU}(N_{\mathrm{c}}^{})$ gauge theories with quark fields
belonging to the fundamental representation. $N_{\mathrm{c}}^{}$ is
considered as a free parameter and the limit of large values of
$N_{\mathrm{c}}^{}$ is taken, while, to ensure a stable limit, the
coupling constant is assumed to scale as
$g\sim 1/N_{\mathrm{c}}^{1/2}$. This limit has been introduced
by 't Hooft, who has studied its general properties
\cite{'tHooft:1973jz,'tHooft:1974hx}. It has been found that in
this limit QCD catches the main properties of confinement, while
being simplified with respect to secondary complications, like quark
pair creation or inelasticities.  $1/N_{\mathrm{c}}^{}$ plays here
the role of a perturbative parameter and allows the classification
of Feynman diagrams according to their topology and relevance
(planar, nonplanar, etc.).
\par
The properties of the theory concerning meson and baryon states
have been analyzed by Witten \cite{Witten:1979kh}. The analysis
is done with the aid of two-point, three-point and four-point
functions of quark color-singlet bilinear operators (currents),
$j(x)=(\overline q\Gamma q)(x)$, where $\Gamma$ represents Dirac
matrices, and the study of their large-$N_{\mathrm{c}}^{}$ behavior.
\par
It is found, from the study of the two-point functions, that, at
large-$N_{\mathrm{c}}^{}$, the hadronic spectrum is saturated by an
infinite number of free stable mesons, made of
a quark, an antiquark and gluons. (The infinite number is dictated
by asymptotic freedom.) Their masses are in general finite:
\be \lb{e2}
M_n^{}=O(N_{\mathrm{c}}^0),\ \ \  n=1,2,\ldots\ .
\ee
Many-meson states contribute only at subleading orders in
$1/N_{\mathrm{c}}^{}$. Baryons, which at large $N_{\mathrm{c}}^{}$
are essentially made of $N_{\mathrm{c}}^{}$ quarks, have masses
that grow like $N_{\mathrm{c}}^{}$.
\par
From the study of the three-point and four-point functions, one
deduces the behavior of the three-meson and four-meson effective
couplings, which are vanishing at large $N_{\mathrm{c}}^{}$
(Fig. \rf{f3}):
\be \lb{e3}
g(MMM)=O(N_{\mathrm{c}}^{-1/2});\ \ \ \ \ 
g(MMMM)=O(N_{\mathrm{c}}^{-1}).
\ee
\bfg 
\bc
\epsfig{file=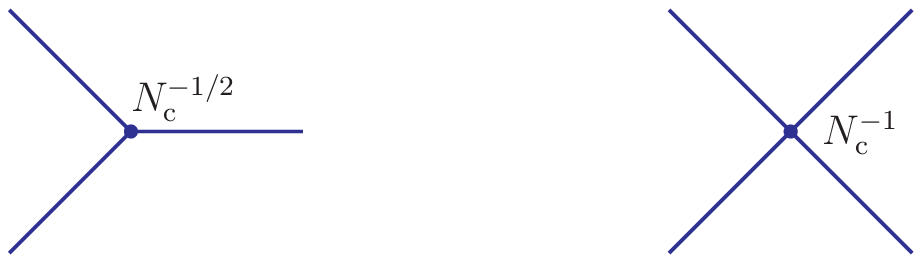,scale=0.75}
\caption{Generic three-meson and four-meson effective couplings.} 
\lb{f3} 
\ec
\efg
These behaviors imply that, at large $N_{\mathrm{c}}^{}$, the mesons
decouple from each other and become free particles.
\par
From Eqs. (\rf{e3}) one deduces the strong decay width behavior
of the mesons:
\be \lb{e4}
\Gamma(M)=O(N_{\mathrm{c}}^{-1}),
\ee
which also expresses their stability property at large
$N_{\mathrm{c}}^{}$.
\par

\section{Tetraquarks at large {\boldmath{$N_{\mathrm{c}}^{}$}}}
\lb{s3}

Can we have similar predictions with tetraquarks? To this end,
one may consider the analog of a bilinear current, a four-quark
color-singlet current, Eq. (\rf{e1}), and its two-point function.
It turns out that, at large $N_{\mathrm{c}}^{}$, the latter
is dominated by disconnected pieces of two-point bilinear
currents:
\be \lb{e5}
\langle T(x)T^{\dagger}(0)\rangle_{\stackrel{{\displaystyle=}}
{N_\mathrm{c}^{}\rightarrow \infty}}
\langle j(x)j^{\dagger}(0)\rangle\ \langle j(x)j^{\dagger}(0)\rangle.
\ee
The right-hand side describes the propagation of two free
ordinary mesons \cite{Coleman:1985}. No tetraquark pole can
appear at this order.
\par
This fact has been considered as a theoretical proof of the
non-existence of tetraquarks as elementary stable particles,
which could survive, like the ordinary mesons, in the
large-$N_{\mathrm{c}}^{}$ limit.
\par
Recently, Weinberg has observed that if tetraquarks exist
as bound states in the large-$N_{\mathrm{c}}^{}$ limit with
finite masses, even if they contribute to subleading diagrams, the
crucial point is the qualitative property of their decay widths: 
are they broad or narrow? In the latter case, they might be 
observable. He has shown that, generally, they should be narrow,
with decay widths of the order of $1/N_{\mathrm{c}}^{}$, which
is compatible with the stability assumption in the 
large-$N_{\mathrm{c}}^{}$ limit \cite{Weinberg:2013cfa}.
\par
Knecht and Peris have shown that in a particular exotic 
channel, tetraquarks should even be narrower, with decay widths
of the order of $1/N_{\mathrm{c}}^{2}$ \cite{Knecht:2013yqa}.
\par
Cohen and Lebed have shown, for more general exotic
channels, with an analysis based on the analyticity properties of
two-meson scattering amplitudes, that the decay widths should
be of the order of $1/N_{\mathrm{c}}^{2}$
\cite{Cohen:2014tga}.
\par
Lucha, Melikhov and Sazdjian show, on the basis of the
singularity analysis of Feynman diagrams with the Landau
equations \cite{Landau:1959fi,Itzykson:1980rh}, that tetraquark
two-meson decay widths of order $1/N_{\mathrm{c}}^{2}$ actually
represent the most general case. For fully exotic tetraquarks
(four different quark flavors) two different tetraquarks, each having
a preferred decay channel, would be needed
\cite{Lucha:2017mof,Lucha:2017gqq}.
\par
Maiani, Polosa and Riquer impose additional selection rules to make
the behaviors compatible with the diquark model and predict two-meson
decay widths of order $1/N_{\mathrm{c}}^{3}$ \cite{Maiani:2016hxw}
and $1/N_{\mathrm{c}}^{4}$ \cite{Maiani:2018pef}.
\par
The particular case of the light meson scattering amplitudes
has been studied by Pel\'aez and Rios within the framework of chiral
perturbation theory \cite{Pelaez:2006nj,Pelaez:2015qba} .
The existence of scalar mesons with a tetraquark type structure has
been found. In the large-$N_{\mathrm{c}}^{}$ limit, their masses
and decay widths diverge as $N_{\mathrm{c}}^{1/2}$. Such behaviors
do not fit the characteristics of compact tetraquarks, as conjectured
by Weinberg. These scalar mesons would rather fit a molecular type 
structure; this is also supported by a direct resolution of the
four-body Bethe-Salpeter equation \cite{Eichmann:2015cra}.
\par
The large-$N_{\mathrm{c}}^{}$ behavior of meson scattering
amplitudes in connection with lattice calculations has also been
studied in \cite{Guo:2013nja}.
\par

\section{Weak point of the large-{\boldmath{$N_{\mathrm{c}}^{}$}}
analysis} \lb{s4}

Contrary to the case of two-point functions of quark bilinear
currents and their saturation by ordinary meson states, possible
tetraquark contributions to two-meson scattering amplitudes are
competing with the background of two-meson states, which consistently
saturate, on qualitative grounds, the corresponding correlation
functions. Unless one does detailed quantitative calculations of
the numerical coefficients of Feynman diagrams and compares them
with the contributions of hadronic intermediate states, the
hypothesis of an eventual absence of tetraquark states does not lead
to any qualitative inconsistency \cite{Lucha:2017gqq}.
\par
Therefore, the predictions made for tetraquark decay amplitudes are
based on the assumption of their possible existence and should be
considered as upper bounds.
\par

\section{Line of approach} \lb{s5}

The study of possibly existing tetraquark properties is done through
the analysis of meson-meson scattering amplitudes \cite{Cohen:2014tga,
Lucha:2017mof,Lucha:2017gqq}.
\par
One considers four-point correlation functions of color-singlet quark
bilinears,
\be \lb{e6}
j_{ab}^{}=\overline q_a^{}q_b^{},
\ee
having a coupling with a meson $M_{ab}^{}$:
\be \lb{e7}
\langle 0|j_{ab}^{}|M_{ab}^{}\rangle = f_{M_{ab}^{}}^{};\ \ \ \ \ 
f_M^{}=O(N_c^{1/2}).
\ee
($a,b$ refer to flavor indices.) Spin and parity are ignored here,
since they are not relevant for the qualitative aspects that are
deduced.
\par
One should consider all possible channels where a tetraquark may be
present.
\par
To be sure that a QCD diagram may contain a tetraquark contribution
through a pole term, one has to check that it receives a four-quark
contribution in its $s$-channel singularities, plus additional gluon
singularities that do not modify the $N_{\mathrm{c}}^{}$-behavior of
the diagram. Their existence is checked with the use of the Landau
equations \cite{Landau:1959fi,Itzykson:1980rh}.
\par
Diagrams that do not have $s$-channel singularities, or have only 
two-particle singularities (quark-antiquark), cannot contribute 
to the formation of tetraquarks at their $N_{\mathrm{c}}^{}$-leading
order. They should not be taken into account for the 
$N_{\mathrm{c}}^{}$-behavior analysis of the tetraquark properties.
\par
We consider here the case of fully exotic tetraquarks, containing
four distinct quark flavors, which we denote by 1,2,3,4, with meson
currents
\be \lb{e8}
j_{12}^{}=\overline q_1^{}q_2^{},\ \ \ \
j_{34}^{}=\overline q_3^{}q_4^{},\ \ \ \
j_{14}^{}=\overline q_1^{}q_4^{},\ \ \ \
j_{32}^{}=\overline q_3^{}q_2^{}.
\ee
The following scattering processes are considered:
\bea
\lb{e9}
& &M_{12}^{}+M_{34}^{}\ \rightarrow\ M_{12}^{}+M_{34}^{},\ \ \ \ \
\mathrm{Direct\ channel\ I};\\
\lb{e10}
& &M_{14}^{}+M_{32}^{}\ \rightarrow\ M_{14}^{}+M_{32}^{},\ \ \ \ \
\mathrm{Direct\ channel\ II};\\
\lb{e11}
& &M_{12}^{}+M_{34}^{}\ \rightarrow\ M_{14}^{}+M_{32}^{},\ \ \ \ \
\mathrm{Recombination\ channel}.
\eea
\par 
The `direct' four-point functions are 
\be \lb{e12}
\Gamma_I^{(\mathrm{dir})}=\langle j_{12}^{}j_{34}^{}j_{34}^{\dagger}
j_{12}^{\dagger}\rangle\ ,\ \ \ \ \ \ 
\Gamma_{II}^{(\mathrm{dir})}=\langle j_{14}^{}j_{32}^{}
j_{32}^{\dagger}j_{14}^{\dagger}\rangle\ .
\ee
Samples of $N_{\mathrm{c}}^{}$-leading and subleading diagrams for
$\Gamma_{I}^{(dir)}$ are presented in Fig. \rf{f4}.
\par
\bfg 
\bc
\epsfig{file=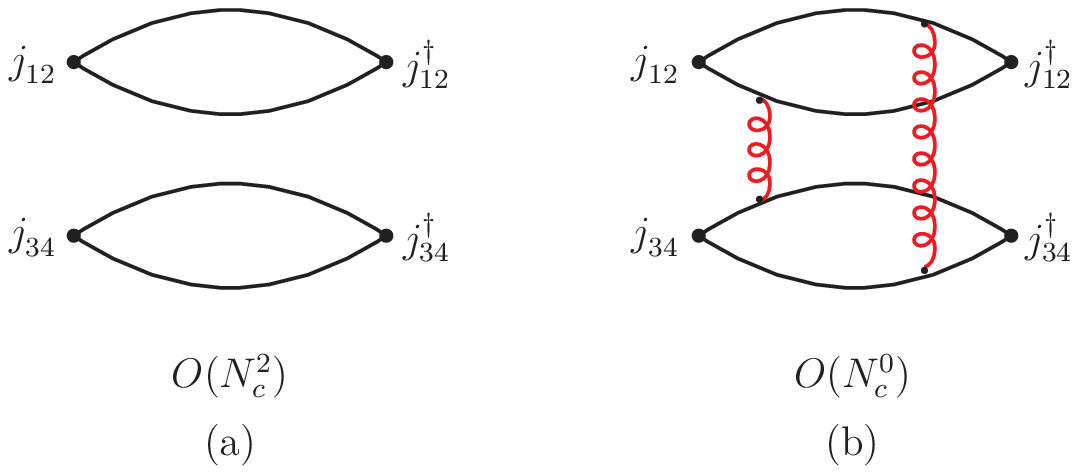,scale=1}
\caption{Samples of leading and subleading diagrams for the direct
channel I.}
\lb{f4}
\ec
\efg
Similar diagrams also exist for $\Gamma_{II}^{(dir)}$.
\par
Only diagram (b) may receive contributions from tetraquark states. 
\par
The `recombination' 4-point function is
\be \lb{e13}
\Gamma^{(\mathrm{recomb})}=\langle j_{12}^{}j_{34}^{}j_{32}^{\dagger}
j_{14}^{\dagger}\rangle\ .
\ee
Samples of leading and subleading diagrams are presented in Fig.
\rf{f5}.
\bfg
\bc
\epsfig{file=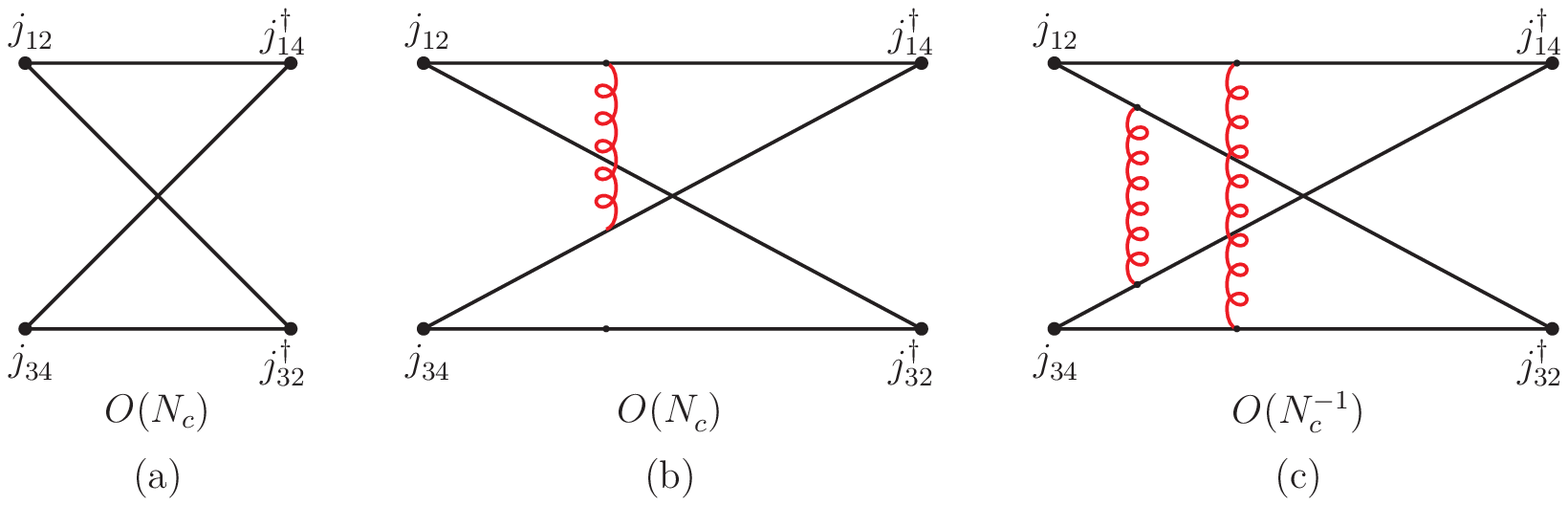,scale=0.7}
\caption{Samples of leading and subleading diagrams for the
recombination channel.}
\lb{f5}
\ec
\efg
Only diagram (c) may receive contributions from tetraquark 
states. Contrary to apparencies, diagrams (a) and (b) do not
have $s$-channel singularities. Their singularities lie
either in the $t$- and $u$-channels, or contribute to the
external meson propagators or vertices.
\par
We note that direct and recombination scattering amplitudes have
different behaviors in $N_{\mathrm{c}}^{}$. Therefore, each
channel imposes a different constraint. Consideration of only
one channel would lead to incomplete solutions.
\par
The solution requires the contribution of two different tetraquarks,
$T_A^{}$ and $T_B^{}$, say, each having different couplings to
the meson pairs. One finds for the tetraquark -- two-meson
transition amplitudes: 
\bea
\lb{e14}
& &A(T_A^{}\rightarrow M_{12}^{}M_{34}^{})=O(N_c^{-1}),\ \ \ \ \ \ 
A(T_A^{}\rightarrow M_{14}^{}M_{32}^{})=O(N_c^{-2}),\\
\lb{e15}
& &A(T_B^{}\rightarrow M_{12}^{}M_{34}^{})=O(N_c^{-2}),\ \ \ \ \ \ 
A(T_B^{}\rightarrow M_{14}^{}M_{32}^{})=O(N_c^{-1}).
\eea
\bfg 
\bc
\epsfig{file=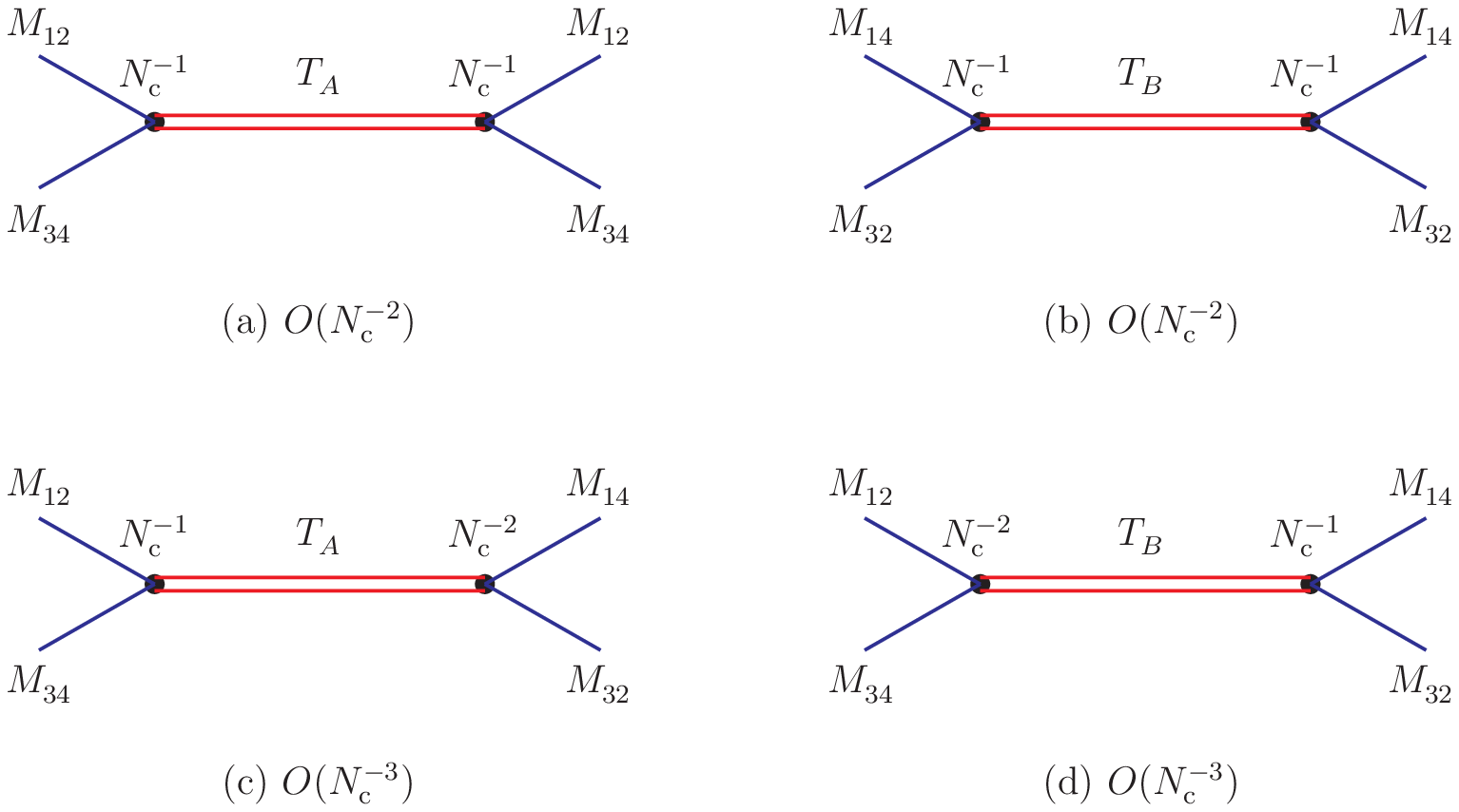,scale=0.75}
\caption{Dominant tetraquark intermediate states in meson-meson
scattering. (a): direct channel I; (b): direct channel II;
(c) and (d): recombination channel.}
\lb{f6}
\ec
\efg
\par
The two-meson decay widths of the teraquarks are
\be \lb{e16}
\Gamma(T_A^{})\sim\Gamma(T_B^{})=O(N_{\mathrm{c}}^{-2}),
\ee
which are much smaller than those of ordinary mesons [Eq. (\rf{e4})].
A diagrammatic representation of the tetraquark intermediate states
with their couplings to the external mesons is given in Fig. \rf{f6}.
\par
There is also a generation of background meson-meson effective 
interaction by means of four-meson effective couplings represented
in Fig. \rf{f7}. Notice that the recombination (quark exchange)
effective coupling dominates; it is actually generated by the
leading diagrams of the recombination channel ((a) and (b) of Fig.
\rf{f5}).
\par
\bfg 
\bc
\epsfig{file=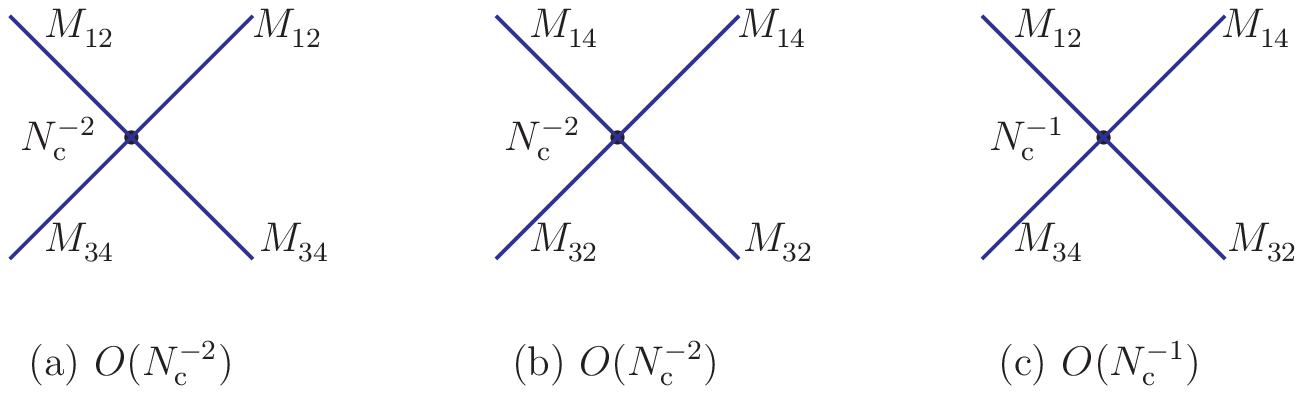,scale=0.75}
\caption{Four-meson effective couplings generated in the direct
channel I (diagram (a)), direct channel II (diagram (b)) and the
recombination channel (diagram (c)).}
\lb{f7}
\ec
\efg
These, in turn, generate meson loops (Fig. \rf{f8}).
\bfg 
\bc
\epsfig{file=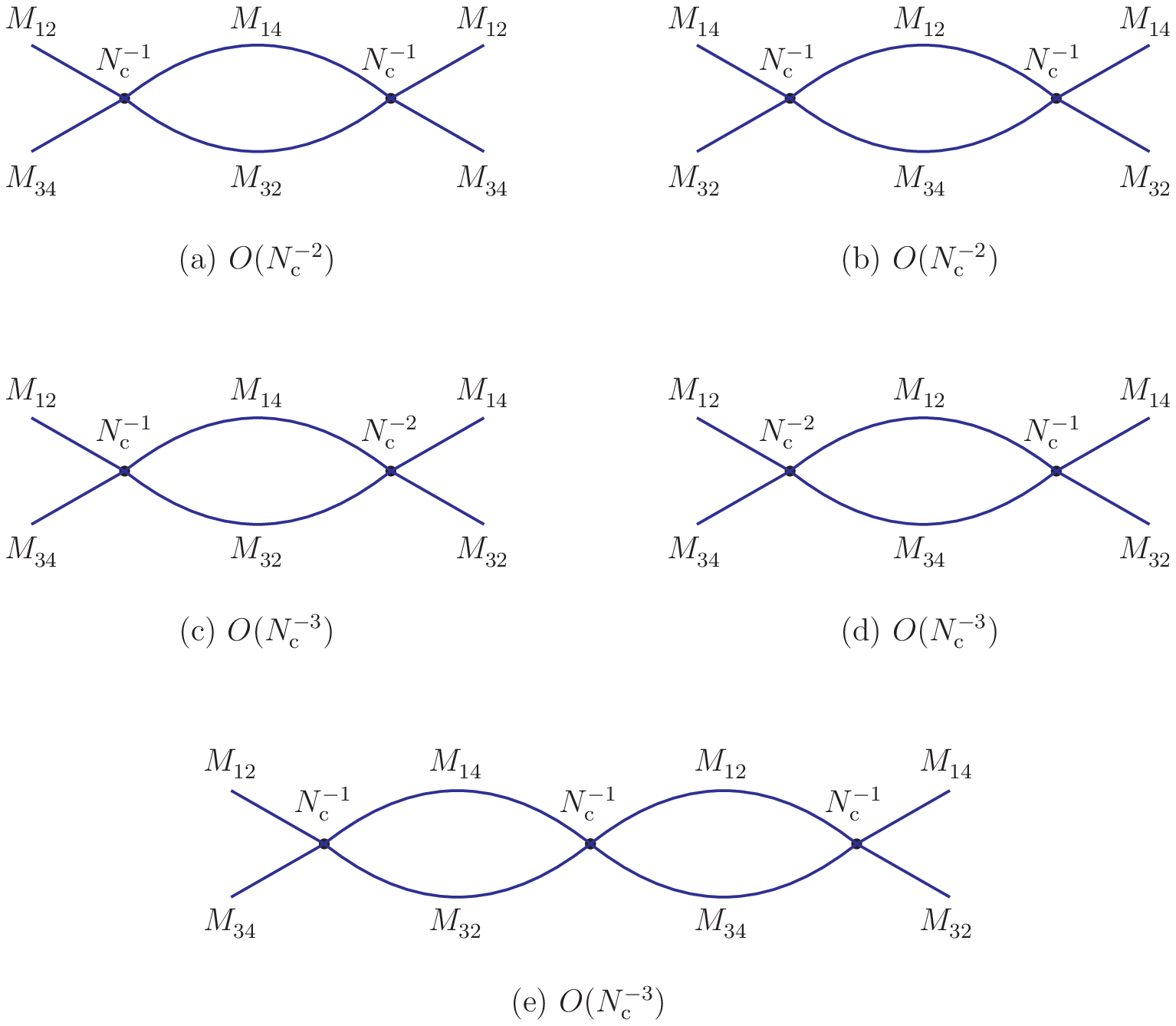,scale=0.6}
\caption{Two-meson intermediate state contributions.}
\lb{f8}
\ec
\efg
\par
From the four-meson couplings of Fig. \rf{f7} and the transition
amplitudes of Eqs. (\rf{e14}) and (\rf{e15}), one can
reconstitute an effective interaction Lagrangian expressed in
terms of quark color-singlet bilinears, from which one deduces
the dominant structure of the two tetraquarks $T_A^{}$ and $T_B^{}$
in terms of the latter quantities \cite{Lucha:2017gqq}. One finds:
\be \lb{e17}
T_A^{}\ \sim\ (\overline q_1^{}q_4^{})(\overline q_3^{}q_2^{}),
\ \ \ \ \ \   
T_B^{}\ \sim\ (\overline q_1^{}q_2^{})(\overline q_3^{}q_4^{}), 
\ee
mixings of order $1/N_{\mathrm{c}}^{}$ between the two
configurations being possible.
\par
The above result favors a color singlet-singlet structure of the
tetraquarks.
\par

\section{Dynamical aspects} \lb{s6}

The fact that we have two different tetraquarks, each having a
structure made of two color-singlet clusters, raises a few
questions. 
\par
First, once the color-singlet clusters are formed inside the four-body
system, their mutual interaction can no longer be confining
(Fig. \rf{f9}). 
\bfg 
\bc
\epsfig{file=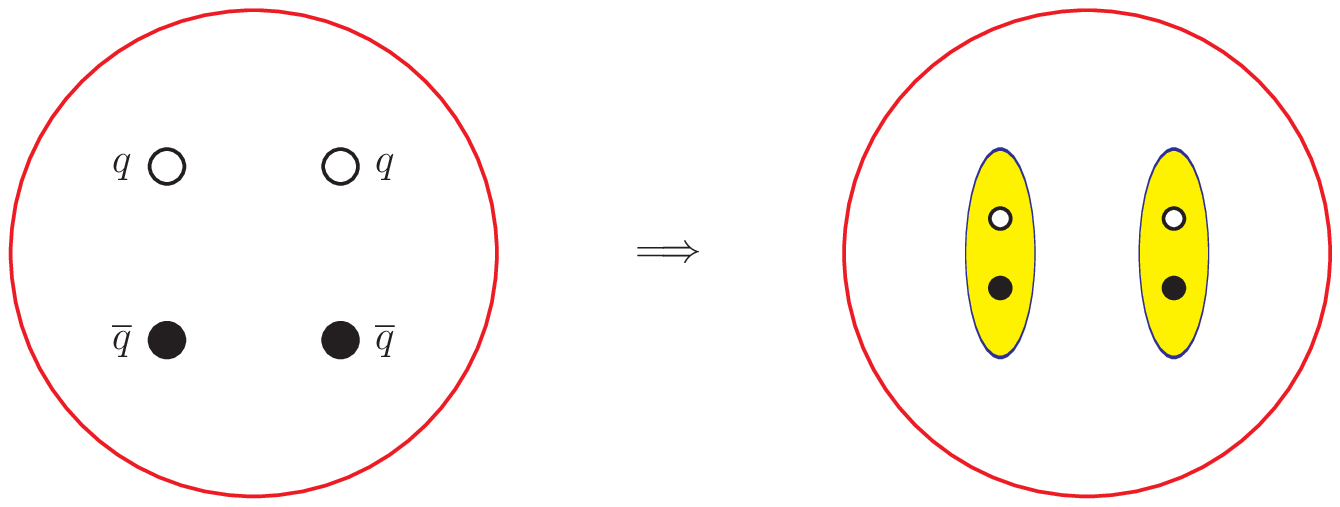,scale=0.5}
\caption{Formation of two color-singlet clusters inside the four-body
  system. (The external circles are a schematic representation of the
  system and do not represent confining bags.)}
\lb{f9}
\ec
\efg
\par
In general, one expects in such a case a short-range interaction
between the two clusters, due to meson exchanges or contact terms,
which would eventually produce a molecular type tetraquark. 
At most, one might expect long-range type Van der Waals forces, 
reminiscent of the confining forces. In that case, the tetraquarks, 
if they exist, would still be loosely bound, similarly to the
molecular type tetraquarks. This possibility has already been
foreseen by Jaffe, stressing that there would not be a clear
phenomenological distinction between the two situations
\cite{Jaffe:2008zz}.
\par
Second, the necessity of having two different tetraquarks to 
accomodate the $N_{\mathrm{c}}^{}$-counting constraints does
not fit the diquark formation scheme, where only one type of diquark 
is expected to be formed, in its color-antisymmetric representation
(Fig. \rf{f10}).
\par
\bfg 
\bc
\epsfig{file=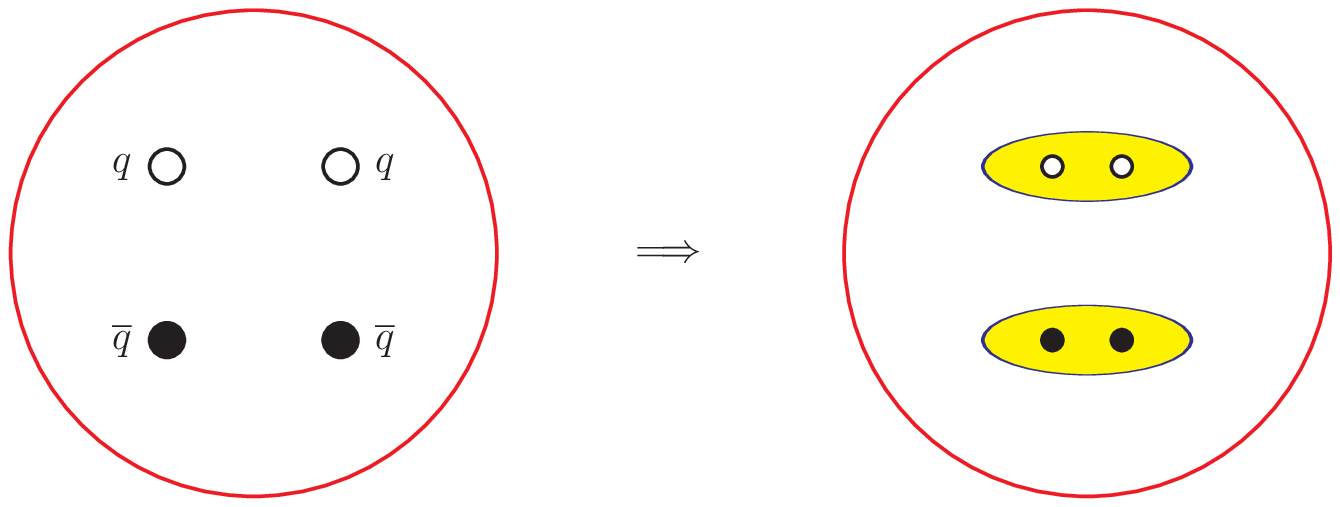,scale=0.5}
\caption{The diquark formation scheme.}
\lb{f10}
\ec
\efg
The binding of the diquark and antidiquark clusters is expected to 
be realized by means of confining forces, hence favoring the
appearance of compact tetraquarks.
\par
To make the diquark scheme compatible with the  
$N_{\mathrm{c}}^{}$-analysis, Maiani, Polosa and Riquer 
impose additional constraints for the selection of diagrams
contributing to the formation of tetraquarks \cite{Maiani:2018pef}.
Only non-planar diagram contributions are retained; this lowers the
contribution of direct channel diagrams by two degrees in
$N_{\mathrm{c}}^{}$. To obtain a consistent solution with one
tetraquark, it is then assumed that, at leading order in
$N_{\mathrm{c}}^{}$, tetraquarks contribute only to the direct
channel diagrams. The decay width into two-mesons is found of the
order of $1/N_{\mathrm{c}}^{4}$.
\par
These constraints, while mathematically correct, require, however,
a more detailed analysis of the corresponding dynamical mechanism
that is at their origin.
\par 

\section{Conclusion} \lb{s7}

The large-$N_{\mathrm{c}}^{}$ limit of QCD allows us to have
a complementary insight into the problem of multiquark states.
Tetraquark and multiquark states, if they exist, do not
generally appear in $N_{\mathrm{c}}^{}$-leading-order terms
and are competing with multimeson background contributions.
The conventional large-$N_{\mathrm{c}}^{}$-based analysis does
not lead to a proof of their existence, but simply gives upper
bounds for their decay or transition amplitudes.
\par
The generic results, in the case of four quark flavors, have the 
tendancy to favor the formation of tetraquarks with two
color-singlet internal clusters. In such a case, the tetraquarks
would probably be loosely bound.
\par
The diquark scheme, which might lead to the emergence of compact 
tetraquarks, requires fine tuning dynamical mechanisms.
\par
The resolution of four-body bound state equations in
conjunction with large-$N_{\mathrm{c}}^{}$ analysis might bring
further information for a better understanding of the question.
\par
\vspace{0.5 cm}

\noindent
\textbf{Acknowledgements.}
I thank the organizers of the Conference Quark Confinement and
the Hadron Spectrum XIII for their invitation to give this talk.
I thank W. Lucha and D. Melikhov for our stimulating
collaborative work on the tetraquark problem. Discussions, during
the conference, with T. Cohen, J. Dudek, M. Knecht and E. Shuryak
are gratefully acknowledged. The figures were drawn with the aid
of the package Axodraw \cite{Vermaseren:1994je}.
\par

\end{document}